\documentclass{PoS}

\title{Towards implications of asymptotically safe gravity for particle physics}

\ShortTitle{
Towards implications of asymptotically safe gravity for particle physics
}

\author{\speaker{Astrid Eichhorn}\thanks{A.~E.~acknowledges support by the Deutsche Forschungsgemeinschaft (DFG) under grant no.~Ei-1037/1 and Danish National Research Foundation under grant DNRF:90. A.~H.~also
  acknowledges support by the German Academic Scholarship
  Foundation.}\\
  CP3-Origins, University of Southern Denmark, Campusvej 55, DK-5230 Odense M, Denmark; Institute for Theoretical Physics, University of Heidelberg,
	Philosophenweg 16, 69120 Heidelberg, Germany\\
        E-mail: \email{eichhorn@sdu.dk}}

\author{Aaron Held\\
       Institute for Theoretical Physics, University of Heidelberg,
	Philosophenweg 16, 69120 Heidelberg, Germany\\
        E-mail: \email{held@thphys.uni-heidelberg.de}}

\abstract{We review aspects of the interplay of asymptotically safe gravity with matter, focusing on the potential predictive power of the quantum scale-symmetry underlying the asymptotically safe fixed point. We explain how an asymptotically safe fixed point for the Standard Model, induced by quantum-gravity fluctuations, might i) render the Standard Model ultraviolet complete and ii)  allow us to calculate the values of some of the Standard-Model couplings. In particular, we highlight that such a fixed point might explain the mass-difference between the top and bottom quark.}

\FullConference{ALPS 2019 An Alpine LHC Physics Summit \\
		April 22 - 27, 2019\\
		Obergurgl, Austria}

\begin{document}

\section{Invitation: A proposal for simplicity}
Given the LHC results up to the date of the writing of these proceedings, the status of the Standard Model as a description of the building blocks of matter is theoretically viable up to the Planck scale. This is a consequence of the mass of the Higgs particle being $m_h=125\, \rm GeV$,
such that neither a Landau pole before the Planck scale, nor an instability of the Higgs potential is realized. The Standard Model faces four challenges: i) Landau poles, signalling a nonperturbative triviality problem, occur for the Abelian hypercharge coupling $g_Y$ and in the Higgs-Yukawa sector beyond the Planck scale; ii) the gravitational interaction is the only one of the known four fundamental  forces that is not included; iii) the perturbatively renormalizable couplings are free parameters, with no way of calculating their values\footnote{This includes the  hierarchy problem, linked to the small value of the Higgs mass in Planck units. Yet, the values of the other couplings, such as the gauge couplings, cannot be calculated either, and remain free parameters.}; iv) the observational evidence for dark matter as well as the matter-antimatter-inequality cannot be explained.
\\
Could the first three of these challenges be connected?  Quantum fluctuations of gravity, becoming significant at Planckian energies, could trigger an ultraviolet (UV) completion of the Standard Model \cite{Eichhorn:2017ylw} with  enhanced predictive power. This might allow us to calculate some of the free parameters of the Standard Model from first principles \cite{ Eichhorn:2017ylw,Shaposhnikov:2009pv,Eichhorn:2017lry,Eichhorn:2018whv}. The proposal is based on the asymptotic-safety paradigm that we briefly review now, see \cite{Reuter:2019byg,Percacci:2017fkn} for reviews.

\section{Scale invariance as a fundamental symmetry of nature}
Scale invariance is (approximately) realized in many instances in the natural world, with fractal-like, self-similar patterns, occurring, e.g., in ferns, clouds and waves. The, perhaps bold, idea of the asymptotic-safety program is that scale invariance could be a much deeper principle of Nature.
Asymptotic safety is a nontrivial realization of scale invariance in a quantum field theory. Generically, quantum field theories are not scale invariant, as quantum fluctuations trigger a scale-dependence of the interactions even in classically scale-invariant models. Yang-Mills theory is  an example, featuring a nonvanishing beta function. Yet, it realizes scale invariance asymptotically, as the UV regime exhibits asymptotic freedom.

Asymptotic safety can be thought of as a generalization of asymptotic freedom: instead of approaching a vanishing value in the UV, the running couplings approach constant finite values, realizing quantum scale invariance and thereby being ``safe'' from Landau-pole like behavior. 

\begin{figure}[t]
\includegraphics[width=0.32\linewidth]{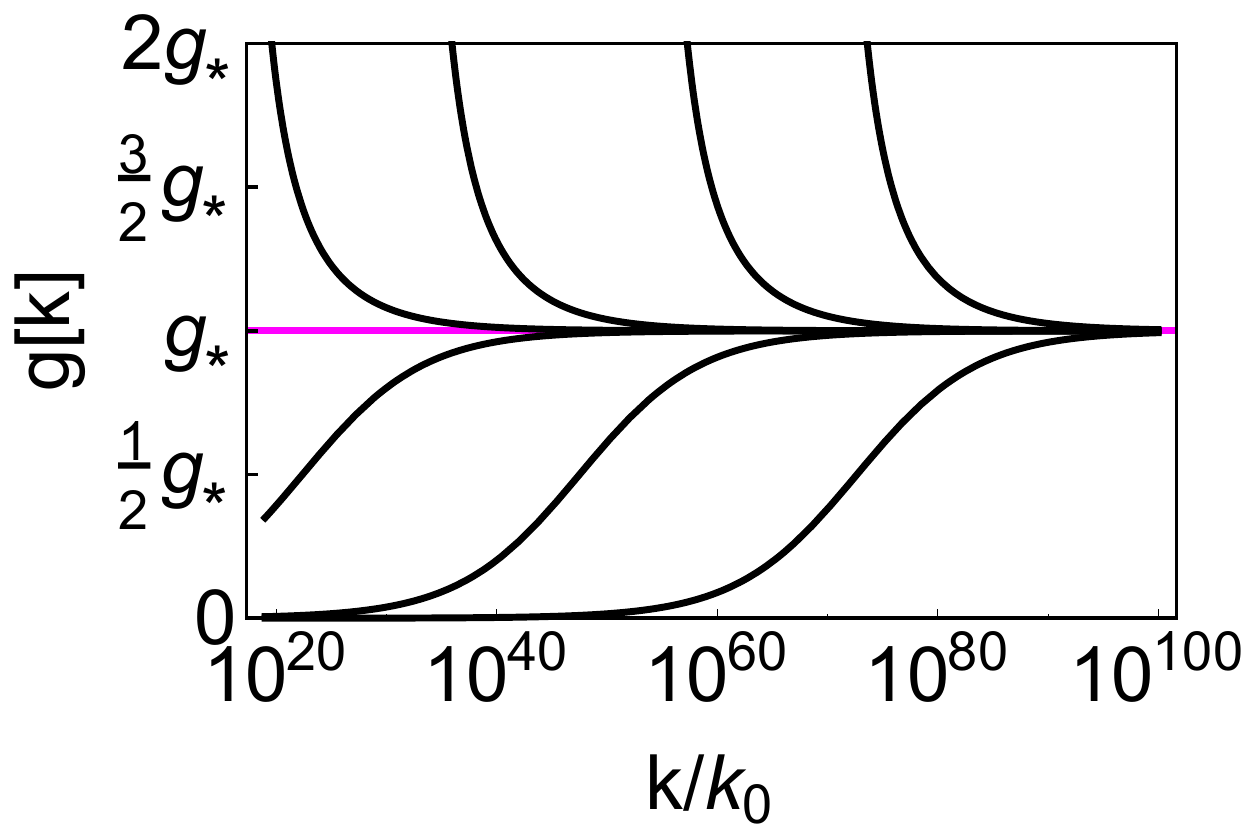}\quad \includegraphics[width=0.32\linewidth]{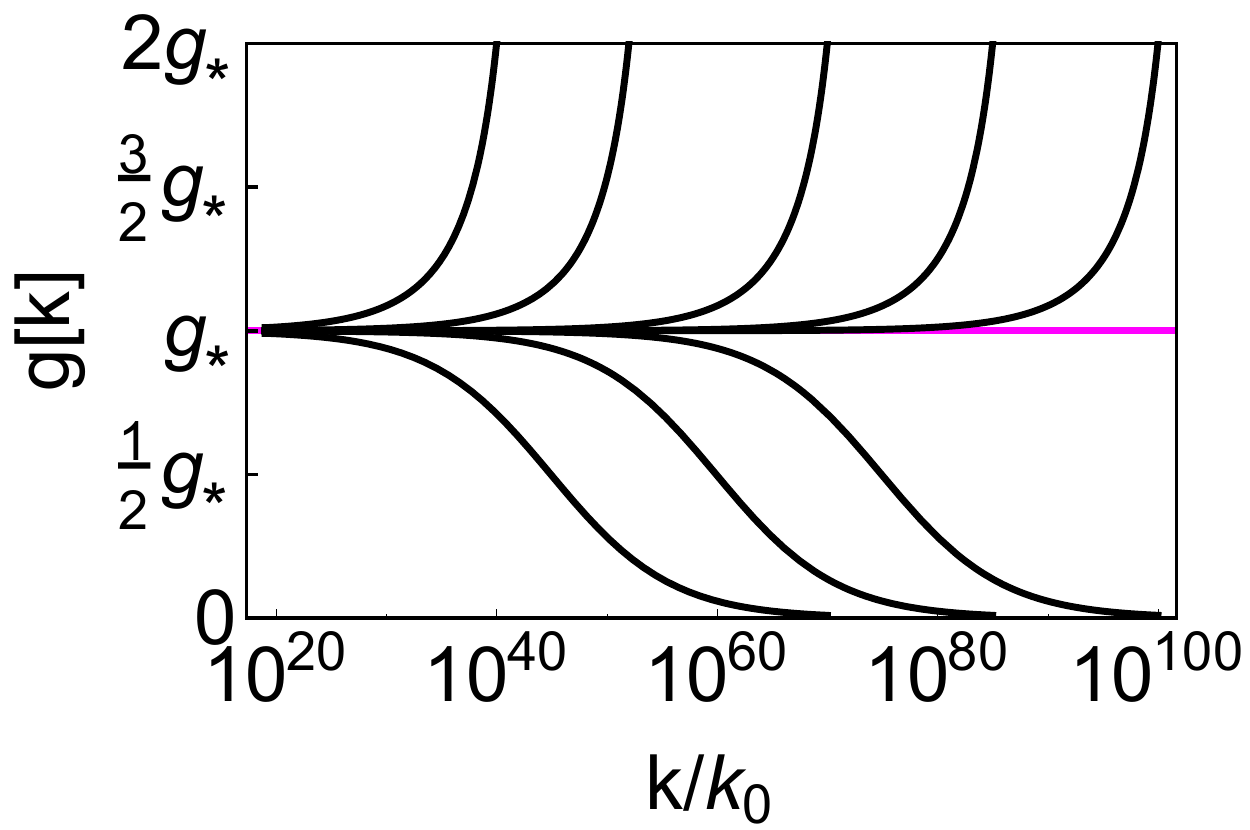}\quad\includegraphics[width=0.25\linewidth]{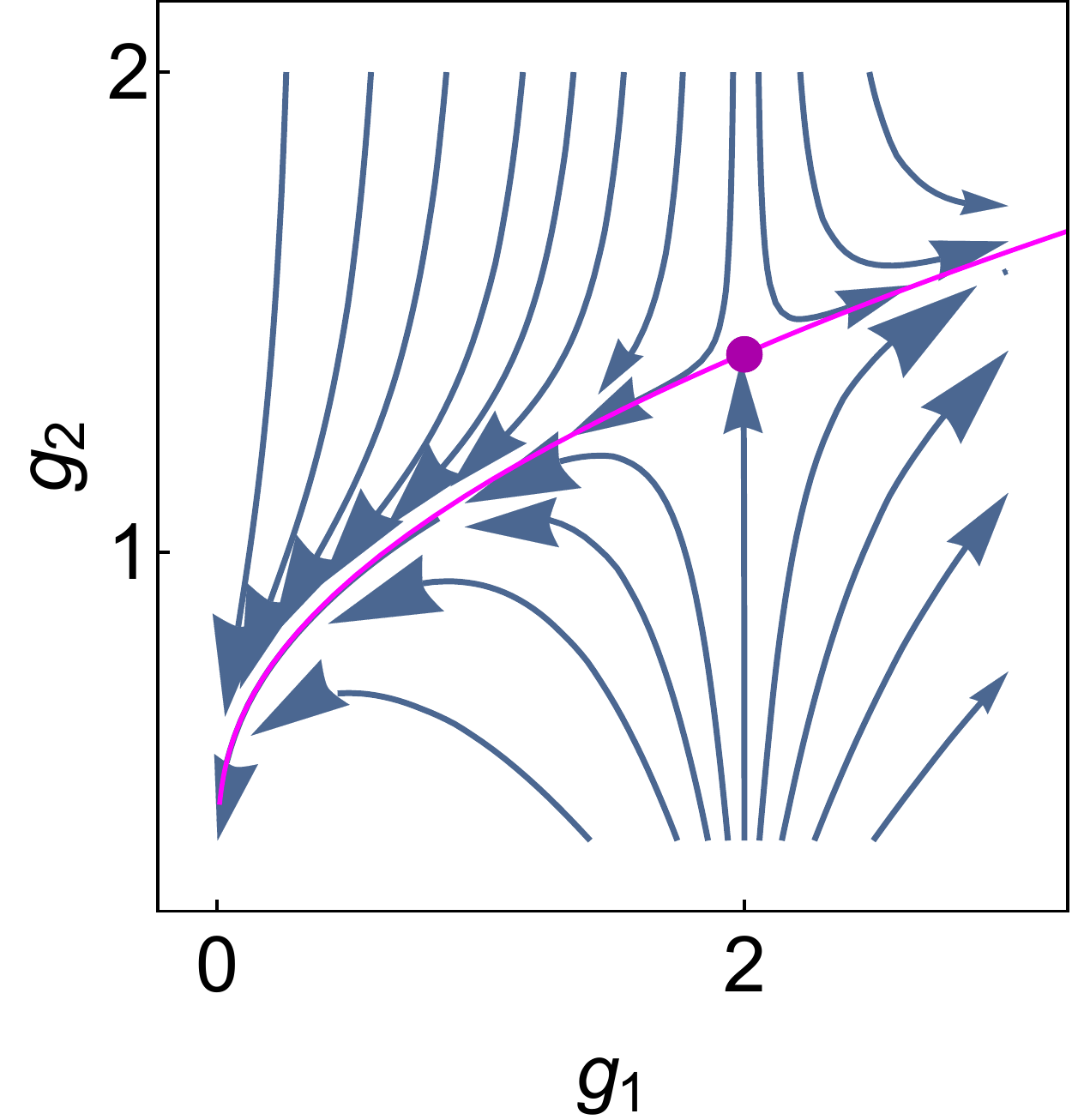}
\caption{\label{fig:relillu} We illustrate IR-repulsive, i.e., relevant (left panel) and IR-attractive, i.e., irrelevant (central panel) directions of a fixed point. The underlying beta functions are given by $\beta_g= \pm g(g-g_{\ast})$, with the negative (positive) sign applying to the left (central) panel. In the right panel we illustrate a UV critical hypersurface (magenta) with curvature in the space of the couplings $g_1, g_2$: On the surface which realizes scale symmetry in the UV, the value of $g_2$ is determined at all scales in terms of $g_1$; fixed-point trajectories only feature one free parameter, namely the value of $g_1$ at a given input scale.}
\end{figure}

Since asymptotic safety entails a symmetry, it gives rise to relations between the interactions that must be satisfied such that the symmetry can be realized. 
The number of remaining free parameters is given by the number of infrared (IR) repulsive directions of the interacting fixed point, the so-called relevant directions. Along those directions, the RG flow can deviate from the scale-invariant fixed-point regime. Thus, the IR value of a relevant coupling encodes the deviation from scale-invariance that characterizes low-energy physics. Conversely, realizing scale-symmetry in the UV does not impose restrictions on the low-energy value of a relevant coupling, cf.~left panel of Fig.~\ref{fig:relillu}.
The predictive power of asymptotic safety lies in fixing the values of IR attractive, i.e., irrelevant couplings \emph{at all scales}. Unlike for the relevant couplings, where asymptotic safety only determines the fixed-point values, but not the low-energy values, irrelevant couplings are fully calculable (in terms of the relevant ones). In the simplest case, quantum fluctuations  force them to remain at their fixed-point value at all scales, cf.~central panel in Fig.~\ref{fig:relillu}. In the more typical case, the UV critical hypersurface -- the hypersurface in which all RG trajectories emanating out of the fixed point lie -- exhibits curvature. This means that the irrelevant couplings change as a function of scale,  but only by following the relevant couplings, cf.~right panel in Fig.~\ref{fig:relillu}.

This discussion holds for all couplings in theory space, i.e., all couplings compatible with the symmetries of the theory, including perturbatively renormalizable and
nonrenormalizable ones. Their beta functions must all vanish at a fixed point. At weakly-interacting fixed points, higher-order (perturbatively nonrenormalizable interactions) typically correspond to irrelevant directions. 

There are several mechanisms generating
asymptotic safety, see e.g., \cite{Eichhorn:2018yfc}  and references therein for an overview, providing a rich playing field for model building. They include a cancellation of one-loop versus higher-loop effects,
(potentially non-perturbative) cancellations between distinct
degrees of freedom, and, the mechanism key for the present setting, a cancellation of canonical scaling versus quantum effects. For example, in
Yang-Mills theory in $d=4+\epsilon$ dimensions, to
leading order in $\epsilon$ the
dimensionless counterpart of the  gauge coupling runs according to:
\begin{equation}
\label{eq:epsilonExp-nAgauge}
\beta_{g}= \epsilon\, g - \beta_0\, g^3+...,
\end{equation}
where $-\beta_0>0$ is the one-loop coefficient in $d=4$. The competition of the antiscreening effect of quantum fluctuations with the positive canonical scaling term leads to an interacting fixed point at
\begin{equation}
g_{\ast} = 
\sqrt{\epsilon / \beta_0}.
\end{equation}
This fixed point is infrared repulsive in the gauge coupling, exemplifying the general pattern that \emph{a model that is asymptotically free in its critical dimension $d_c$ exhibits an IR repulsive interacting fixed point\footnote{It is an intriguing question what the respective upper critical dimension of such a fixed point is. As it becomes more strongly coupled away from $d_c$, this question is in general difficult to answer.} in $d=d_c+\epsilon$}. Conversely, \emph{a model that is trivial in its critical dimension $d_c$ exhibits an IR attractive fixed point in $d=d_c-\epsilon$.} This can lead to a very interesting situation in models in which different interactions feature distinct critical dimensionalities. In particular, the critical dimension for the Newton coupling is $d=2$, and studies of the $\epsilon$ expansion indeed indicate asymptotic safety in $d=2+\epsilon$, with functional RG techniques \cite{Wetterich:1992yh,Balog:2019rrg} providing compelling hints for asymptotic safety in $d=4$, based on the seminal work by Reuter \cite{Reuter:1996cp}. The gravitational contribution to matter beta functions evaluated with functional RG techniques is generically linear in the respective matter couplings, thus acting akin to an effective change in dimensionality. This  could be of particular 
importance for marginally irrelevant Standard Model couplings, such as the Abelian gauge coupling and the Yukawa couplings: 
An effective dimensional reduction
generates an IR attractive interacting fixed point in those matter couplings. 
This fixed point is a Planck-scale phenomenon, 
but
its imprints could persist to the electroweak scale. This link between the Planck scale and the electroweak scale is worth elucidating further.

\section{A link that matters}
The standard lore is that physics at different scales decouples, and that one does not need knowledge of the microphysics in order to describe the macrophysics. Most of the microscopic information is ``washed out'' by the RG flow. For instance, in perturbation theory, this pertains to higher-order interactions, which are quickly drawn to tiny values by the RG flow towards the IR, irrespective of their microscopic value. This appears to remain true for higher-order interactions in the asymptotically safe gravity-matter setting: Higher-order matter interactions, such as, e.g., four-fermion interactions \cite{Eichhorn:2011pc} 
are present at the asymptotically safe fixed point\footnote{We stress that all results on the effect of asymptotically safe quantum gravity are obtained within truncations of the functional RG within a systematic expansion, and are thus subject to systematic errors.}. Yet, they are irrelevant (as they would be in a canonical power-counting), and thus the RG flow below the Planck scale quickly drives 
them
towards zero.
Yet, there is 
no
complete decoupling of physics at different scales: 
The effective, macroscopic description of the system always features free parameters, which are determined by the microphysics. As an illustrative example, the viscosity in hydrodynamic descriptions of fluids is a free parameter of the effective description. Its value is actually determined by the microscopic interactions of the molecules making up the fluid in question. Although the viscosity originates at molecular scales, it can be measured at much larger scales -- allowing certain microscopic descriptions which yield the wrong value of the viscosity to be ruled out experimentally by measurements at much larger scales than those of the molecular interactions.

In the case of the Standard Model, those parameters that are sensitive to the microphysics are the couplings that run logarithmically below the Planck scale (as well as the perturbatively relevant ones). Due to their slow running, they can keep a ``memory'' of the UV initial conditions. The key point in an asymptotically safe setting is that some of these couplings could be irrelevant and thus \emph{calculable at the Planck scale}. This provides a unique initial condition for the RG flow in the Standard Model below the Planck scale, which maps the initial condition to an IR value that can be compared to values extracted from experiments.

We highlight that exploring the possibility to derive Standard-Model couplings from quantum gravity is interesting from two perspectives: Firstly, it could 
explain some of the seemingly arbitrary constants in the Standard Model. Secondly, it could make aspects of quantum gravity testable, as it links physics at the Planck scale to physics at the electroweak scale. Let us stress that the link works most directly if there are no new-physics scales between the electroweak scale and the Planck scale but can be extended to cases with new physics.

We point out the non-universality of the gravitational contribution, which is a simple consequence of the nonvanishing dimensionality of the Newton coupling. In the Standard Model beta functions without gravity, non-universality sets in at three loops, with gravity at one loop, requiring us to keep track of the scheme. Here, we report on results obtained with functional RG techniques. We stress that these require a Euclidean setting, instead of the Lorentzian one.

\section{Predictive power of an enhanced symmetry}
\label{sec:predictivePower}

The 1-loop contribution to the running of the Abelian hypercharge coupling is screening. Hence, in the Standard Model without gravity, it develops a transplanckian triviality problem. Its 1-loop beta function, including a gravitational contribution $f_g$, reads
\begin{equation}
    \beta_{g_Y} = - f_g\, g_Y + {\frac{41}{6\cdot 16 \pi^2}}\; g_Y^3\;.\label{eq:betagy}
\end{equation}
Approximations for $f_g$, obtained in systematic truncations of the functional Renormalization Group, cf.~\cite{Daum:2009dn, Harst:2011zx, Folkerts:2011jz,Eichhorn:2017lry,Christiansen:2017cxa} and references therein, find $f_g\geqslant0$. If it is non-vanishing, the gravitational contribution therefore is antiscreening. This resembles a dimensional reduction, cf.~Eq.~\ref{eq:epsilonExp-nAgauge}, of the ``effective'' spacetime probed by gauge fluctuations. This could be understood as yet another form of dynamical dimensional reduction in quantum gravity, cf.~\cite{Carlip:2017eud}. \\
In the beta function Eq.~\ref{eq:betagy}, competing screening (matter) and antiscreening (gravitational) quantum fluctuations lead to a transplanckian flow governed by two fixed points,
\begin{equation}
    g_{Y\,\ast\rm AF}^2 = 0\;,
    \quad\quad\quad\quad\quad\quad\quad\quad\quad
    g_{Y\,\ast\rm AS}^2 ={\frac{6\cdot 16 \pi^2}{41}} f_g \;.
\end{equation}
\begin{figure}[t]
\centering
\includegraphics[width=.6\textwidth]{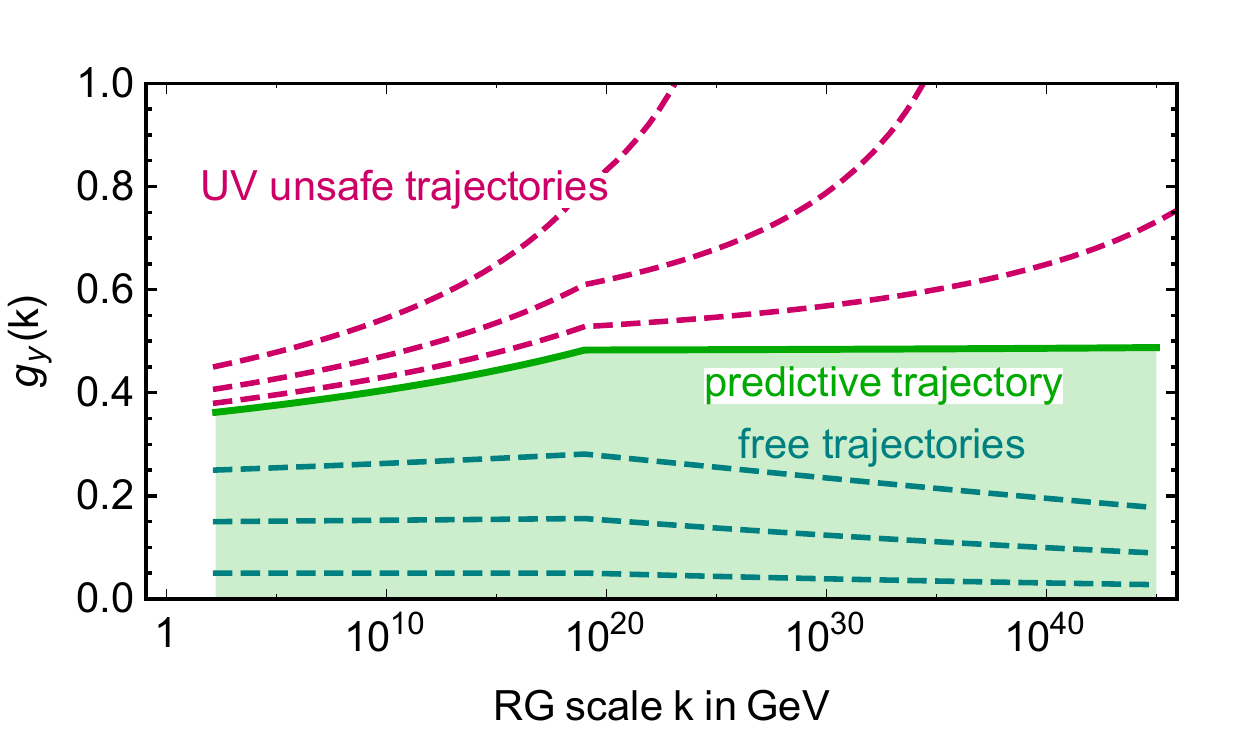}
\caption{Gravitationally induced UV-completions of the Abelian gauge coupling $g_Y$, governed by the asymptotically free ($\ast\rm AF$) and the IR-attractive, asymptotically safe fixed point ($\ast\rm AS$), shown for $f_g=10^{-2}$, see also \cite{Eichhorn:2017lry}.}
\label{fig:upperBound-U1}
\end{figure}
As one confirms by linearizing the flow around each fixed point, $\ast\rm AF$ is IR-repulsive, while $\ast\rm AS$ is IR-attractive. Gravitational fluctuations could thereby tame the triviality problem \cite{Harst:2011zx, Eichhorn:2017lry}. 
Quantitatively, this UV-completion becomes available whenever the Planck-scale value of the gauge coupling is lower than its IR-attractive interacting fixed point, i.e., $g_{Y\,\rm Planck}^2 < g_{Y\,\ast\rm AS}^2$ \cite{Eichhorn:2017ylw, Eichhorn:2017lry}. For larger values of the gauge coupling, screening contributions remain dominant and a Landau pole, although shifted to higher scales, persists in the flow described by Eq.~\ref{eq:betagy}.
\\
$\ast\rm AS$ offers a UV-completion with enhanced predictivity, as it realizes a precise balance of gravitational and matter fluctuations. Perturbations to larger (smaller) values are driven back to $g_{Y\,\ast\rm AS}^2$ by matter (gravitational) fluctuations and dynamically restore scale symmetry towards the IR, cf.~Fig.~\ref{fig:upperBound-U1}. Scale symmetry at $\ast\rm AS$ thus connects a single IR-value of $g_Y$ to the magnitude of asymptotically safe gravitational fluctuations encoded in $f_g$. The value of $f_g$ depends on the fixed-point values of gravitational couplings. Calculating these with sufficient accuracy provides an observational test of the quantum-gravity model.

\section{Why is the top quark such a heavyweight?}
The top quark is more than $30$ times heavier than any other observed fundamental fermion. Its $\mathcal{O}(1)$ Yukawa coupling to the Higgs results in a top quark mass of the order of the Higgs vacuum expectation value. 
But why is its Yukawa coupling so much larger than those of the other quarks?

We will argue that its interplay with gravity and the Abelian gauge interaction could be at the heart of the answer.
Any answer which includes gravity has to first explain why gravitational fluctuations do not break chiral symmetry, thereby causing all fermions to acquire Planck-scale, i.e., $\mathcal{O}(10^{19}\,{\rm GeV})$, masses to begin with. Thus, it is crucial for asymptotically safe quantum gravity to preserve the chiral structure of the Standard Model. As has been explicitly found in Euclidean functional RG calculations (subject to systematic truncation uncertainties) \cite{Eichhorn:2011pc}, gravitational fluctuations do not induce Planck-scale fermion condensates, indicating that fermion masses can remain of the order of the electroweak scale, even in the presence of quantum gravity.

\begin{figure}[t]
\centering
\includegraphics[width=.48\textwidth]{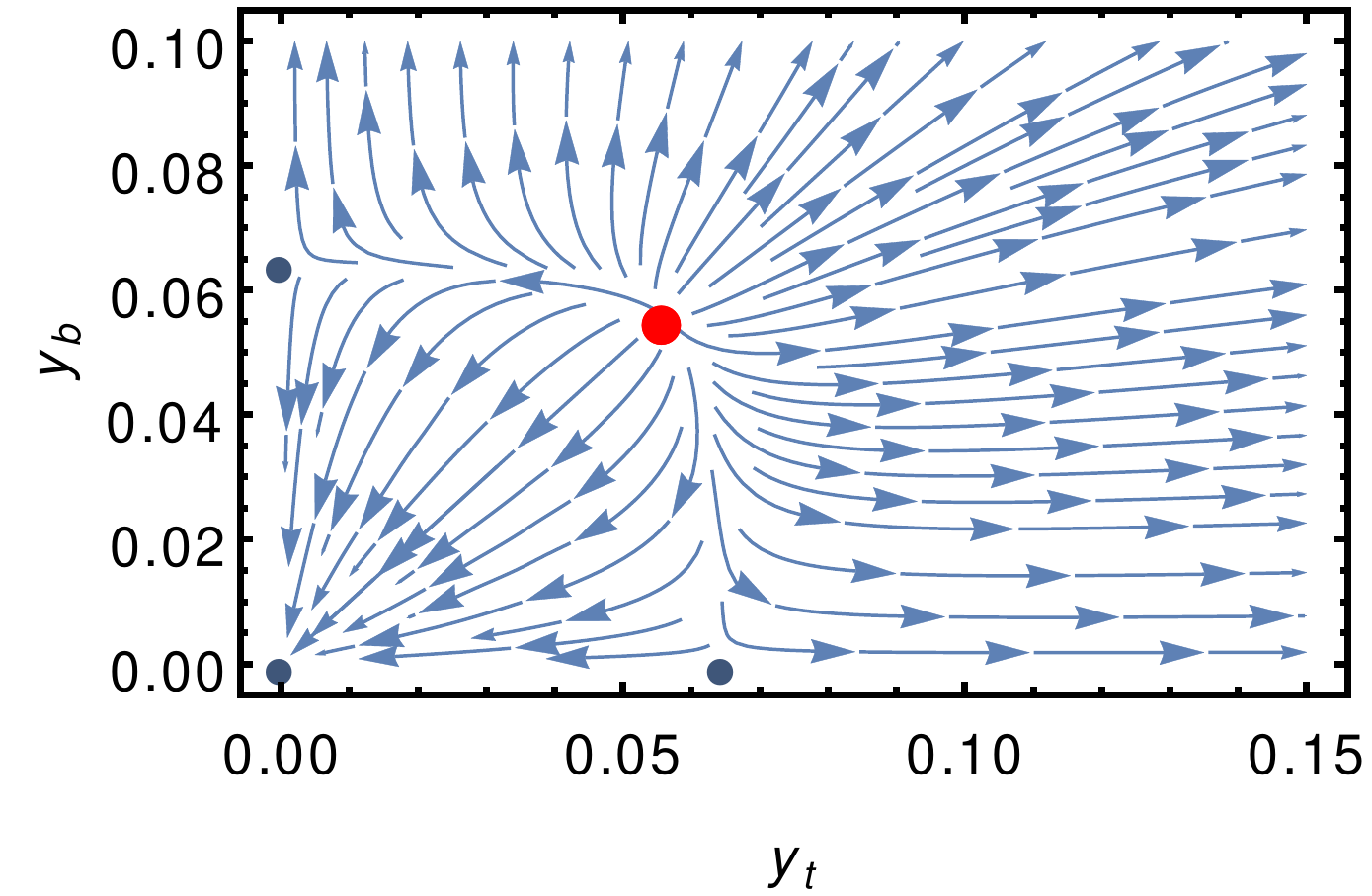}
\hfill
\includegraphics[width=.48\textwidth]{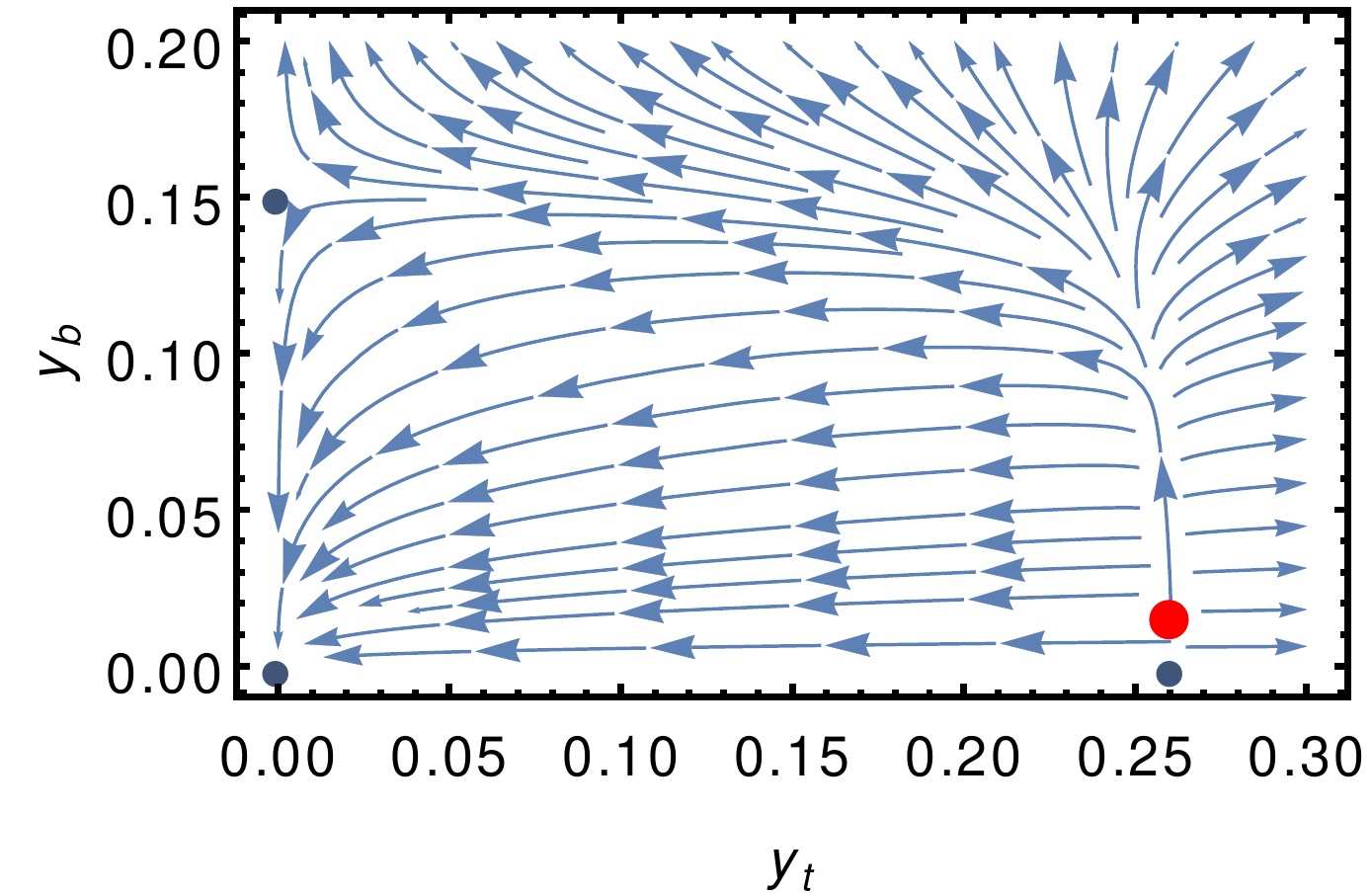}
\caption{Transplanckian flow lines (arrows pointing towards the IR) and fixed point structure of Yukawa couplings in absence (left panel) and presence (right panel) of Abelian gauge interactions. We present the flows, cf.~Eq.~\ref{eq:betaYtb} for $f_y=1.188\times10^{-4}$ as well as $g_Y=0$ (left panel) and $g_Y=0.45$.
Large red and small blue points indicate the respective IR-attractive and the less predictive fixed points, respectively.
The red fixed point realizes the fixed-point relation Eq.~\ref{eq:fp-relation} for $g_{Y, \, \ast}=0$ (left panel) and $g_{Y\, \ast}\neq 0$ (right panel).}
\label{fig:stream-withAndWithoutU1}
\end{figure}

A potential
explanation for the difference between top and bottom mass
may then arise from an interplay between the Higgs, the fermions and gravity, because the logarithmic running of Yukawa couplings conserves an imprint of gravitationally induced quantum scale-invariance. Gravitational UV-completions for Yukawa couplings \cite{Eichhorn:2017ylw, Eichhorn:2018whv} could become available along the same lines as the one for the Abelian hypercharge coupling, cf.~Sec.~\ref{sec:predictivePower}. Leading-order gravitational fluctuations, encoded in $f_y$  and evaluated with functional RG techniques, appear in the 1-loop $\beta$-function  for the top (and bottom) Yukawa couplings $y_{t(b)}$ at transplanckian scales, i.e.,
\begin{eqnarray}
\label{eq:betaYtb}
    \beta_{y_{t \,(b)}} &=
\frac{y_{t \,(b)}}{16\,\pi^2}\;\left(\frac{3 y_{b \,(t)}^2}{2}+ \frac{9 y_{t \,(b)}^2}{2} \right) - f_y\, y_{t \,(b)}  -\frac{3 y_{t \,(b)}}{16\,\pi^2}\left(
Y_Q^2+Y_{t \,(b)}^2\right)g_Y^2\;,
\end{eqnarray}
where the first, second and third term arise from Higgs-fermion, gravitational and Abelian gauge fluctuations, respectively. Since they are of no importance to the following discussion, we have omitted 1-loop contributions from non-Abelian gauge couplings (included in \cite{Eichhorn:2018whv}) as well as those of other (negligibly small) Yukawas. By $Y_Q$, $Y_t$, and $Y_b$ we denote the hypercharges of the left-handed quark doublet, the right-handed top, and the right-handed bottom singlet.  Neglecting the gauge-contributions for a moment, the fixed-point structure for Yukawa couplings is invariant under a $t\leftrightarrow b$ exchange symmetry, cf.~left-hand panel of Fig.~\ref{fig:stream-withAndWithoutU1}. The most predictive version of quantum scale-invariance would thus result in $M_b\approx M_t$ and offers no explanation for a heavy top. 
Due to the different hypercharges $Y_{t(b)}$ of the (right-handed) top and bottom quark, this symmetry is broken by a non-vanishing gauge contribution, as is present at the IR-attractive fixed point in Sec~\ref{sec:predictivePower}, cf.~Fig.~\ref{fig:upperBound-U1}. As can be checked immediately, a fixed point in Eqs~\ref{eq:betaYtb} and \ref{eq:betagy} at which the Abelian gauge coupling $g_Y$, the top Yukawa $y_t$ and the bottom Yukawa $y_b$ are all non-vanishing, cf.~right panel in Fig.~\ref{fig:stream-withAndWithoutU1}, has to obey the fixed-point relation
\begin{equation}
    \label{eq:fp-relation}
    y_{t\,\ast}^2 - y_{b\,\ast}^2 = \frac{1}{3}g_{Y\,\ast}^2\;.
\end{equation}
Here, we have specialized to Standard Model charges $Y_Q=1/6$, $Y_t=2/3$, and $Y_b=-1/3$. This relation is independent of the gravitational contributions $f_g$ and $f_y$. It predicts a smaller bottom than top Yukawa coupling, i.e., $y_{b\,\ast}^2 < y_{t\,\ast}^2$, as a generic consequence of quantum scale-invariance. This hierarchy can persist towards the IR, cf.~\cite{Eichhorn:2018whv}, where it results in $M_b< M_t$.

\begin{figure}[t]
\centering
\includegraphics[width=.47\textwidth]{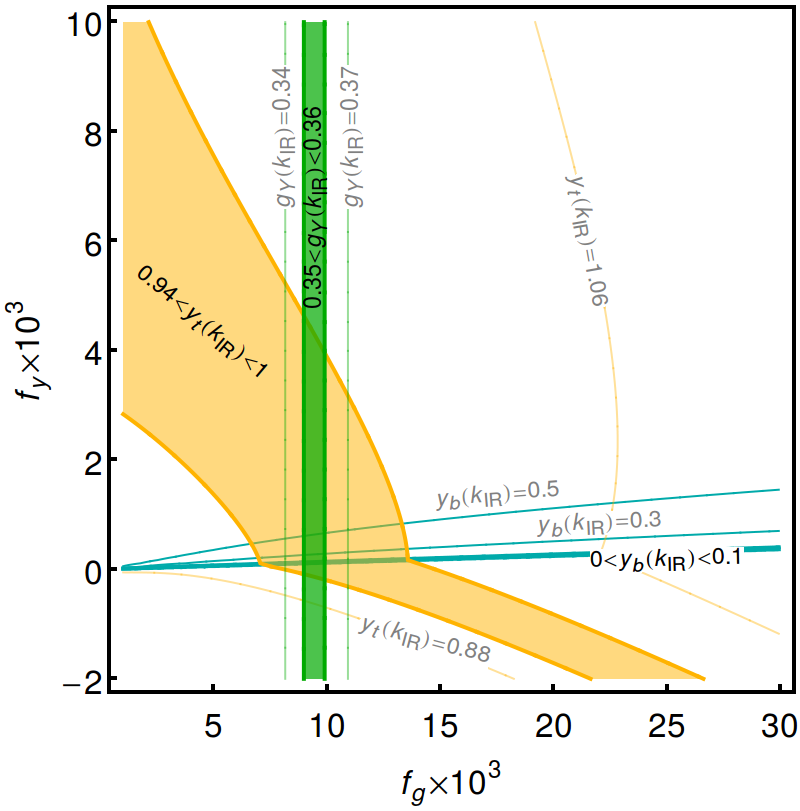}
\hfill
\includegraphics[width=0.47\linewidth]{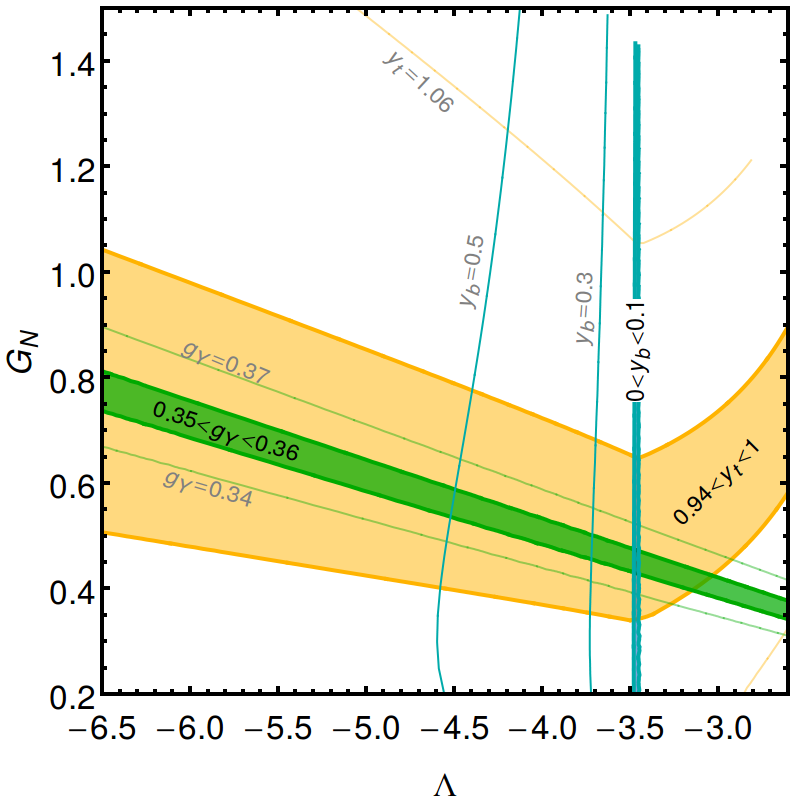}
\caption{Left panel: Bands of IR values of the couplings $g_{Y,\,{\rm IR}}$, $y_{t,\,{\rm IR}}$ and $y_{b,\,{\rm IR}}$ at the top-mass scale $k_{\rm IR} = 173\,{\rm GeV}$, as a function of asymptotically safe gravity contributions $f_g$ and $f_y$, see also \cite{Eichhorn:2018whv}.
Right panel: The same bands in the plane spanned by the dimensionless Newton coupling $G_{N}$ and the dimensionless cosmological constant $\Lambda$, cf.~\cite{Eichhorn:2017ylw, Eichhorn:2017lry}.}
\label{fig:fyfgPlot}
\end{figure}

To discuss whether this proposed mechanism could work quantitatively, we point out that the UV-critical hypersurface of the fixed point, cf.~Eq.~\ref{eq:fp-relation}, fixes all three couplings $g_{Y,\,{\rm IR}}$, $y_{t,\,{\rm IR}}$ and $y_{b,\,{\rm IR}}$ in the IR. In Fig.~\ref{fig:fyfgPlot}, we highlight IR values for these couplings in the plane of possible values for the two gravitational contributions $f_g$ and $f_y$. Clearly, the existence of an intersection of the corresponding bands is non-trivial. Put differently, one can fix the a priori unknown gravitational fluctuations by, for instance, the experimental values of $g_y$ and the bottom mass, i.e., $f_g(g_{Y} (k_{\rm IR}=173\, {\rm GeV}))$ and $f_y(y_{b} (k_{\rm IR}=173\, {\rm GeV}))$. Then, a heavy top, i.e., $y_{t}  (k_{\rm IR}=173\, {\rm GeV})\approx 1$, follows automatically. After electroweak symmetry breaking, this corresponds to a top mass of roughly $M_t\approx 180\;{\rm GeV}$, subject to substantial systematic uncertainties due to truncations of the functional RG equation.\\
We emphasize that quantitative agreement with observation requires (i) new physics contributions $f_g$ and $f_y$ in the vicinity of the Planck scale, (ii) a universal gravitational contribution $f_g$ to all (Abelian and non-Abelian) gauge couplings and (iii) the 
quark charges 
$Y_t=2/3$ and $Y_b=-1/3$ of the Standard Model, see \cite{Eichhorn:2018whv} for more details.

To test the
predictive power of the asymptotically safe UV completion, it is necessary to obtain sufficiently converged approximations of the microscopic (fixed-point) values of  gravity couplings which contribute to $f_g$ and $f_y$. In particular, it is a crucial question, whether the highly predictive region in Fig.~\ref{fig:fyfgPlot} can be accessed.
First approximations for $f_g$ \cite{Daum:2009dn, Harst:2011zx, Folkerts:2011jz,Eichhorn:2017lry,Christiansen:2017cxa} and $f_y$ \cite{Zanusso:2009bs,Oda:2015sma,Eichhorn:2016esv, Eichhorn:2017eht} include contributions of the Newton coupling $G_N$ and the cosmological constant $\Lambda$, cf.~right panel of Fig.~\ref{fig:fyfgPlot}. The latter can act as an effective mass-like suppression, potentially rendering gravitational fluctuations perturbative and resulting in $f_g$ and $f_y$ of the approximately right order of magnitude. First results, including the feedback of Standard-Model matter fluctuations, yield indications for such a suppression by a negative cosmological constant \cite{Dona:2013qba}, see also \cite{Meibohm:2015twa,Biemans:2017zca,Wetterich:2019zdo}, but substantial systematic uncertainties related to truncations of the RG flow are present.
Moreover, as gravity enters a regime of quantum scale invariance, an infinite tower of higher-order couplings will also contribute. For Yukawa couplings, and with the caveat that any finite-order truncation of these functions will induce artificial unitarity violations, these contributions have been studied up to the curvature-squared level in \cite{Eichhorn:2017eht}.  Induced matter couplings feed back into the Yukawa couplings at sub-leading order  \cite{Eichhorn:2017eht}.

There are hints that $d=4$ could be the critical dimension for asymptotic safety \cite{Eichhorn:2019yzm} based on the interplay of quantum gravity and matter as discussed here. An asymptotically safe fixed point in pure gravity appears to exist also in other dimensions \cite{Fischer:2006fz}. For the combined gravity-matter system, going significantly beyond $d=4$ appears to be impossible if one demands a solution to the Abelian triviality problem: Due to the negative canonical dimension of the gauge coupling in $d>4$, the gravitational contribution has to be strongly nonperturbative. This appears to trigger novel divergences in higher-order matter interactions. Accordingly, $d=4$ might be the only dimensionality in which gravity can induce a consistent UV completion for matter.

\section{Outlook}
We have reviewed recent results  
which hint 
at
a potential quantum-gravity induced UV-completion of the Standard Model. This potential regime of quantum scale-invariance could offer a mechanism to solve the Landau-pole problem while simultaneously increasing the predictive power. In particular, it could thereby offer a microscopic explanation for the fine-structure constant as well as the top and bottom mass. 
 Comprehensive tests of the viability of these ideas require
convergence of the actual gravitational fixed point in the presence of matter, as well as more complete studies of UV completions for the full Standard Model, which remain to be explored in future work.

Coming back to the 4th challenge mentioned in the introduction, it is of interest to understand the restrictions placed by asymptotic safety on dark matter. For instance, the Higgs portal coupling to uncharged scalar dark matter might vanish as a consequence of scale symmetry \cite{Eichhorn:2017als}.

\end{document}